\documentclass[aps,prl,twocolumn,amsthm,amsmath,amssymb, superscriptaddress]{revtex4-2}
\usepackage{graphicx}% Include figure files
\graphicspath{{Figures/}}
\usepackage{subfigure}
\usepackage{epsfig}
\usepackage{ulem}
\usepackage{dcolumn}% Align table columns on decimal point
\usepackage{bm}% bold math
\usepackage{hyperref}% add hypertext capabilities
\hypersetup{colorlinks=true, citecolor=blue, urlcolor=blue, linkcolor=blue}
\usepackage{appendix}
\usepackage{mathrsfs}

\newcommand{\s}{{\sigma}}

\def\beq{\begin{equation}}
\def\eeq{\end{equation}}
\def\bald{\begin{aligned}}
\def\eald{\end{aligned}}
\def\bea{\begin{eqnarray}}
\def\eea{\end{eqnarray}}

\def\ket#1{\left|#1\right\rangle}
\def\avg#1{\left\langle#1\right\rangle}

\renewcommand{\(}{\left(}
\renewcommand{\)}{\right)}
\renewcommand{\[}{\left[}
\renewcommand{\]}{\right]}

\def\Eq#1{Eq.~(\ref{#1})}
\def\Fig#1{Fig.~\ref{#1}}

\begin{document}
\title{Fermi surface symmetric mass generation: a quantum Monte-Carlo study }
\author{Wei-Xuan Chang}
\thanks{These authors contributed equally to the work.}
\affiliation{Beijing National Laboratory for Condensed Matter Physics and Institute of Physics, Chinese Academy of Sciences, Beijing 100190, China}
\author{Sibo Guo}
\thanks{These authors contributed equally to the work.}
\affiliation{Beijing National Laboratory for Condensed Matter Physics and Institute of Physics, Chinese Academy of Sciences, Beijing 100190, China}
\author{Yi-Zhuang You}
\email{yzyou@ucsd.edu}
\affiliation{Department of Physics, University of California, San Diego, CA 92093, USA}
\author{Zi-Xiang Li}
\email{zixiangli@iphy.ac.cn}
\affiliation{Beijing National Laboratory for Condensed Matter Physics and Institute of Physics, Chinese Academy of Sciences, Beijing 100190, China}
\affiliation{University of Chinese Academy of Sciences, Beijing 100049, China}

\begin{abstract}
The symmetric mass generation (SMG) phase is an insulator in which a single-particle gap is intrinsically opened by the interaction, without involving symmetry spontaneously breaking or topological order. Here,  we perform unbiased quantum Monte-Carlo simulation and systematically investigate a bilayer fermionic model hosting Fermi surface SMG in the strongly interacting regime. With increasing interaction strength, the model undergoes a quantum phase transition from an exciton insulator to an SMG phase, belonging to the (2+1)-dimensional O(4) universality class. We access the spectral properties of the SMG phase, resembling a Mott insulating phase with relatively flat dispersion and pronounced spectral broadening. The dispersion of Green's function zeros is extracted from spectral function, featuring a surface at zero frequency precisely located at the original non-interacting Fermi surface, which constitutes a hallmark of the Fermi surface SMG phase. The bilayer model we study is potentially relevant to the newly discovered high-$T_c$ superconductor $\rm{La}_3 \rm{Ni}_2 \rm{O}_7$. Our results in SMG phase qualitatively capture the salient features of spectral function unveiled in recent ARPES experiments, shedding new insight on the underlying physics of $\rm{La}_3 \rm{Ni}_2 \rm{O}_7$.    
\end{abstract}
\date{\today}

\maketitle
{\it Introduction}: Symmetric mass generation (SMG) is a mechanism by which a metal acquires excitation gaps without breaking symmetry or developing topological order\cite{You2022Review}. In recent years, SMG has attracted increasing attention in various areas, including particle physics\cite{BenTov2015JHEP,Wen2013CPL,You2015PRB,You2022PRL,Chandrasekharan2016PRD,Chandrasekharan2017PRD,Wang2020PRR,Tong2022JHEP} and condensed matter physics. In condensed matter physics, SMG is investigated in the framework of interacting topological insulator or superconductor\cite{Fidkowski2010PRB,Shinsei2012PRB,Qi2013NJP,Yao2013PRB,Wang2014Science,Wang2014PRB,Xu2015PRB,Cheng2018PRB,Xu2021arXiv}, particularly the interaction reduced classification of symmetry protected topological phases. Additionally, it is demonstrated that SMG transition point in the Dirac fermions constitutes a fermionic version of deconfined quantum critical point (QCP)\cite{You2018PRX,You2018PRB,You2023PRB2,He2016PRB}, enormously enriching the understanding of exotic QCP beyond Landau's paradigm\cite{Senthil2004Science,Senthil2004PRB,Levin-Senthil2004,Sandvik2007DQCP,Shao2016DQCP,Li2017NC}. 

\begin{figure}[t]
\includegraphics[width=0.4\textwidth]{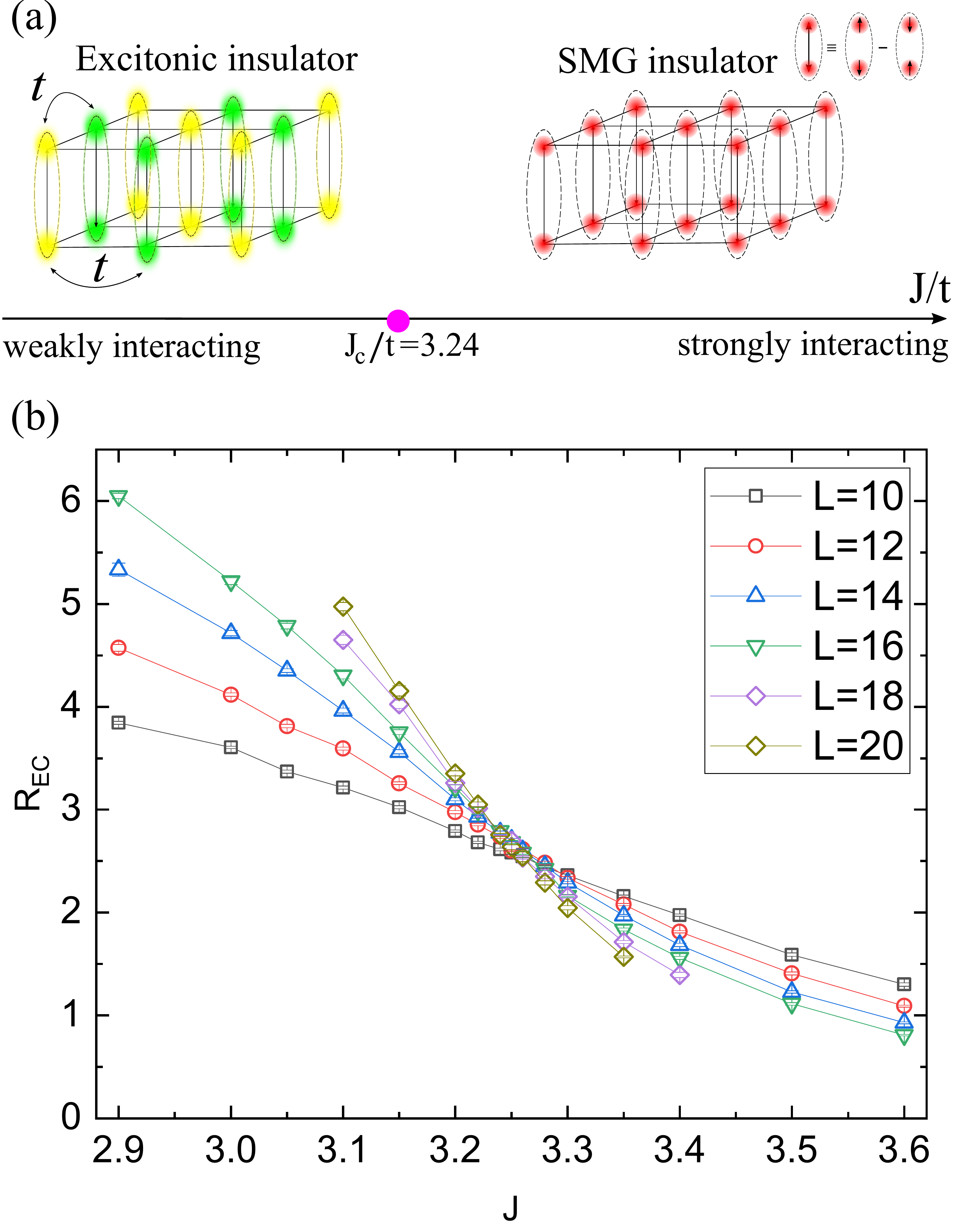}
\caption{ (a) The ground-state phase diagram of \Eq{Hamiltonian}. The system undergoes a quantum phase transition from exciton insulator to SMG phase at $J=3.24$. For the exciton insulator, the yellow (green) bonds denote inter-layer exciton order with momentum $(\pi,\pi)$. For the SMG phase, the inter-layer electrons form a spin singlet, preserving all the lattice and spin rotational symmetries. (b) The results of correlation-length ratio for exciton order as a function of $J$ for different system sizes. }
\label{fig1}
\end{figure}

More recently, Fermi surface SMG\cite{You2023PRB3,You2023arXiv,You2023arXivAnomaly}, denoting the gap opening without SSB or topological order in the Fermi surface, emerges as a fascinating subject due to its fundamental interests relevant to Fermi surface anomaly\cite{Dominic2021PRX,Dominic2021PRL} and Luttinger's theorem\cite{Luttinger1960}.  The bilayer model featuring strong anti-ferromagnetic (AFM) inter-layer spin interaction is proposed as a simple microscopic model to host the Fermi surface SMG phase. Intriguingly, from field theory analysis and perturbative numerical calculation, it is suggested that the poles of Green's function in the pristine gapless fermion state are replaced by the Green's function zeros, which encloses a Fermi volume fulfilling the requirement of Luttinger's theorem\cite{You2023arXiv}. The surface of Green's function zeros at low energy is the hallmark of Fermi surface SMG.  Nevertheless, the unbiased demonstration is still lacking, and it is immensely desirable to investigate Fermi surface SMG by unbiased theoretical approach, unraveling the property of Green's function zeros and other prominent features relevant to experimental observables. 

From the perspective of realistic materials, the bilayer model hosting SMG insulator is potentially relevant to the newly discovered high-$T_c$ superconductor $\rm{La}_3 \rm{Ni}_2 \rm{O}_7$, which exhibits superconductivity with $T_c$ up to 80K under pressure\cite{Wang2023Nature,Cheng2023arXiv2,Cheng2023arXiv,Yuan2023arXiv}. $\rm{La}_3 \rm{Ni}_2 \rm{O}_7$ superconductor harbors a layered structure, with each unit cell containing two layers of $\rm{Ni}\rm{O}_2$. The DFT calculation\cite{Yao2023PRL,Hirschfeld2023arXiv,Zhang2023arXiv,Kuroki2023arXiv,Hu2023arXiv,Zhang2023arXiv3} and ARPES measurements \cite{Zhou2023arXiv} reveal that the low-energy band structure of $\rm{La}_3 \rm{Ni}_2 \rm{O}_7$ under pressure is dominated by the electrons in Ni-$3d_{z^2}$ and Ni-$3d_{x^2-y^2}$ orbitals. With increasing pressure, the material undergoes a structure transition, rendering a strong enhancement of inter-layer AFM Heisenberg interaction between Ni-$3d_{z^2}$ electrons. Distinct from Cuprates superconductors\cite{KeimerReview,Wen2006RMP,ReviewSachdev2003}, although more experimental efforts are required, the magnetic orders have not been detected so far in $\rm{La}_3 \rm{Ni}_2 \rm{O}_7$. Hence, $\rm{La}_3 \rm{Ni}_2 \rm{O}_7$ is a prospective platform for experimental realizations of Fermi surface SMG. Moreover, the theoretical understanding of the underlying mechanism of high-$T_c$ SC remains under development\cite{Jiang2023arXiv,Yao2023arXiv,Wu2023arXiv,You2023arXiv2,Yang2023PRB,Yang2023arXiv,Qin2023arXiv,Werner2023arXiv,Leonov2023PRB,Zhang2023arXiv2,Su2023arXiv,Si2023arXiv,
Yang2023arXiv2,Lu2023arXiv,You2023arXiv3,schlomer2023arXiv,Wu2023arXiv2,Li2023arXiv4}.  
Fathoming the accurate properties of the bilayer model with strong inter-layer spin interaction and the emergent SMG phase by an intrinsically unbiased theoretical approach offers a promising route to demystify the microscopic origin of high-$T_c$ SC in $\rm{La}_3 \rm{Ni}_2 \rm{O}_7$.

To address the aforementioned questions, we perform numerically-exact quantum Monte-Carlo (QMC) simulation\cite{BSS,Hirsch1981PRL,Hirsch1985PRB,Assaadnote,ZXLiQMCreview} to systematically investigate a bilayer fermionic model with inter-layer AFM Heisenberg interaction on square lattice. Remarkably, the model is free from the notorious sign problem\cite{Loh1990PRB,Troyer2005PRL,CJWu2005PRB,Berg2012Science,ZXLi2015PRB,ZXLi2016PRL,LWang2015PRL,TXiang2016PRL,ZXLiQMCreview,Assaadnote}, so the ground-state properties of the model with large system size are accessible. 
We achieve the ground-state phase diagram of the model. In the weak coupling regime, the ground state possesses inter-layer exciton order, degenerated with inter-layer spin-singlet pairing order. With increasing interaction strength, the exciton order is suppressed and the system undergoes a quantum phase transition to a SMG phase without any SSB. The large-scale QMC simulation reveals the transition belongs to the $(2+1)$-dimensional O(4) universality class. Furthermore, we investigate the spectral properties of the SMG phase. The dispersion of the spectral peak resembles of Mott insulator, featuring a relatively flat band and pronounced broadening. The surface of Green's function zeros at zero frequency is extracted from the results of the spectral function, which resides on the original Fermi surface in the non-interacting limit, fulfilling the requirement of Luttinger's theorem. Lastly, we discuss the implication of our numerical results on the $\rm{La}_3 \rm{Ni}_2 \rm{O}_7$ high-$T_c$ superconductors and potential detection of Fermi surface SMG phase in future experiments.

{\it Model}: We consider the bilayer model on a square lattice with inter-layer AFM Heisenberg interaction, as described by the following Hamiltonian\cite{Zhai2009A0905.1711,Congjun2022PRB,Zhang2020F2001.09159,Zhang2020D2006.01140,Nikolaenko2021S2103.05009,Bohrdt2021E2107.08043,Bohrdt2022S2108.04118,Hirthe2023M2203.10027,You2023arXiv}:
\bea\label{Hamiltonian}
&&H= -t\sum_{\avg{ij}\s\alpha}(c^\dagger_{i\s\alpha}c_{j\s\alpha}+\mathrm{h.c.}) + J\sum_{i} \vec{S}_{i,1}\cdot \vec{S}_{i,2},~~~
\eea
where $c^\dagger_{i\s}$ creates an electron on site $i$ with spin polarization $\s=\uparrow,\downarrow$ and layer index $\alpha=1,2$. 
Here $t$ is the hopping amplitude of electrons between nearest-neighbor sites and $J$ denotes the strength of inter-layer AFM 
Heisenberg interaction. Hereafter, we set $t=1$ as the energy unit and focus on the model at half-filling. Remarkably, the model is free from the notorious sign problem such that we can access the ground-state properties of the model with large system sizes through the projective version of QMC simulation\cite{Assaad2001PRB}. The details of the projector QMC algorithm are included in the Supplementary Materials. 

The model preserves total spin SU(2) symmetry. Additionally, the model at half-filling respects pseudospin SU(2) symmetry for each layer\cite{You2023arXiv}. In the non-interacting limit, namely $J=0$, the model at half-filling features a Fermi surface with nesting momentum $(\pi,\pi)$. In the weak coupling regime, the instability with spontaneous symmetry breaking appears due to the Fermi surface nesting. Conversely, in the strong coupling limit $ J \rightarrow \infty$, the ground state wave function is $\ket{\psi} = \prod_i (c^\dagger_{i1\uparrow} c^\dagger_{i2\downarrow}-c^\dagger_{i1\downarrow} c^\dagger_{i2\uparrow}) \ket{0}$. The state is the product of the inter-layer spin singlet state on each site, respecting all symmetries, including the lattice symmetry, the spin rotational symmetry, and the charge conservation symmetries for each layer. In the following, we perform a sign-problem-free QMC simulation to study the accurate ground-state phase diagram of the model for generic coupling $J$. 

\begin{figure}[tb]
\includegraphics[width=0.5\textwidth]{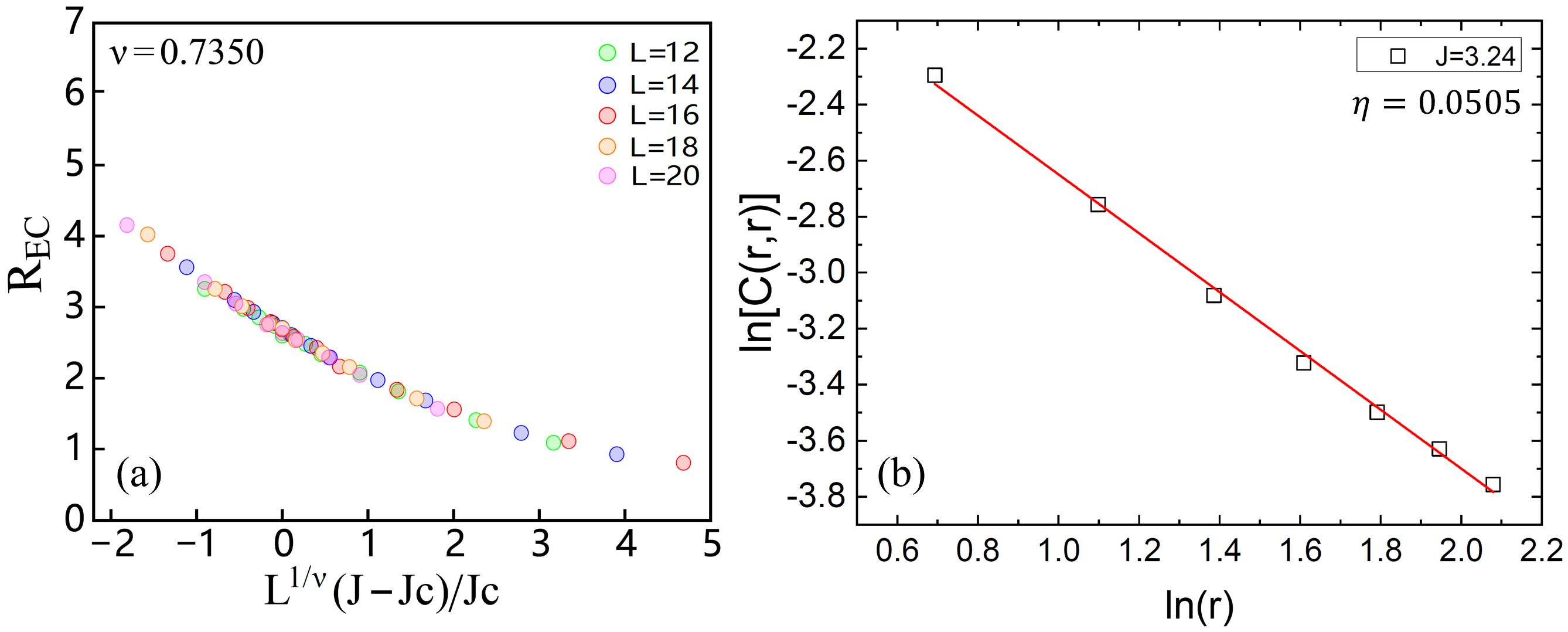}
\caption{(a) The data collapse analysis of correlation-length ratio for excitonic order. The resulting correlation-length exponent $\nu=0.735\pm0.024$. The rescaled correlation-length ratios for different system sizes are collapsed into a single curve. (b) Logarithm of correlation function for exciton order versus $\ln(r)$ at QCP $J=3.24$, giving rise to the anomalous dimension $\eta =  0.051\pm0.036$ from the scaling function $C(r,r) \sim \frac{1}{r^{1+\eta}}$.  }
\label{fig2}
\end{figure}

\begin{figure*}[tb]
\includegraphics[width=0.9\textwidth]{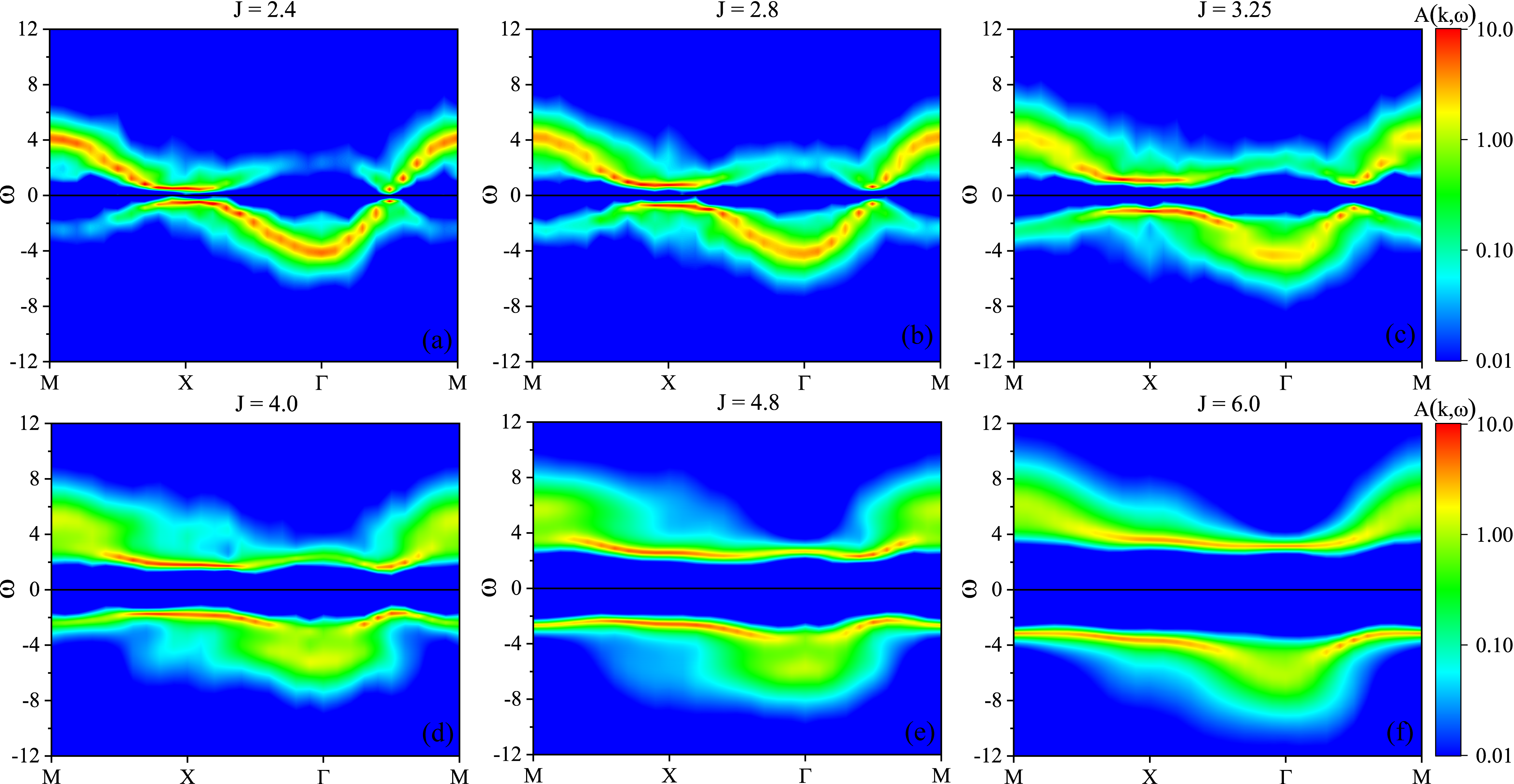}
\caption{The results of spectral function $A(k,\omega)$ along the high-symmetry momentum line for (a) $J=2.4$, (b) $J=2.8$, (c) $J=3.25$, (d) $J=4.0$, (e) $J=4.8$, (f) $J=6.0$.  }
\label{fig3}
\end{figure*}

{\it Ground-state phase diagram}:
We investigate the ground-state properties of the model \Eq{Hamiltonian} with varying interaction strength $J$. To unveil the possible symmetry-breaking phase, we calculate the structure factors of distinct orderings and the corresponding correlation-length ratios, the detailed descriptions of which are included in the Supplementary Materials. The results of structure factors and correlation-length ratios, as included in SM, unambiguously reveal that the inter-layer exciton condensation (EC) is the dominant one in the weak coupling regime. The order parameter of the EC order is given by $\phi_{\rm EC} = \frac{1}{N}\sum_{i\sigma} c^\dagger_{i\sigma1}c_{i\sigma2}(-)^i$ with $(-)^i=\pm1$ for $i\in\text{A,B}$ sublattices, which carries momentum $(\pi,\pi)$ due to Fermi surface nesting. As verified by the scaling analysis of structure factor and correlation-length ratios, the ground state possesses EC long-range order when $J$ is small.

Further increasing interaction strength $J$, the exciton order is suppressed and vanishes at a critical $J$. To identify the quantum critical point $J_c$, we investigate the behaviors of correlation length ratios for different system sizes. The results are depicted in \Fig{fig1}(b), unequivocally revealing a quantum phase transition between EC long-range ordered and disordered phases occurring at $J_c=3.24$. In the parameter regime $J>3.24$, we confirm the absence of SSB in the ground state by evaluating the correlation-length ratio for different orderings, as shown in SM. Moreover, the single-particle gaps are finite in the whole regime of $J$ (as shown in SM), substantiating that the model \Eq{Hamiltonian} hosts Fermi surface SMG insulating phase when $J>3.24$. Hence, we arrive at the ground-state phase diagram as shown in \Fig{fig1}(a).

{\it Quantum criticality}: As aforementioned, in addition to the total spin SU(2) symmetry, the model at half filling also respects the $\rm{SU}(2)_1 \times \rm{SU}(2)_2 $ pesudospin symmetry, where $\rm{SU}(2)_l$ (for each layer $l=1,2$) is generated by $\bm{K}_l=\sum_{i}\frac{1}{2}(-)^i \tilde{c}_{il}^\dagger \bm{\sigma} \tilde{c}_{il}$ with $\tilde{c}_{il}=(c_{il\uparrow},c_{il\downarrow}^\dagger)^\intercal$ being the Nambu spinor. Owing to the enlarged pseudospin symmetry, EC order is degenerate with inter-layer singlet SC order. The degeneracy is explicitly explained by the fact that order parameter of EC $\phi_{\rm EC} = \frac{1}{N}\sum_{i\sigma} c^\dagger_{i\sigma1}c_{i\sigma2}(-1)^i$ is related to the inter-layer SC order $\phi_{\rm SC} = \frac{1}{N}\sum_{i\sigma} (-)^\sigma c^\dagger_{i\sigma1}c^\dagger_{i\bar{\sigma}2}$ by the particle-hole transformation $c_{i\sigma2}\rightarrow (-)^i(-)^\sigma c^\dagger_{i\bar{\sigma}2} $ (a subgroup of $\rm{SU}(2)_1 \times \rm{SU}(2)_2$), where $(-)^\sigma=\pm1$ for $\sigma=\uparrow,\downarrow$ and $\bar{\sigma}$ represents the opposite spin of $\sigma$. EC and inter-layer SC order preserve total spin rotational symmetry, but break pesudospin symmetries $\rm{SU}(2)_1 \times \rm{SU}(2)_2 $, which is equivalent to $\rm{SO}(4)$ symmetry. Consequently, the quantum phase transition from exciton insulator/SC to SMG phase is presumably an O(4) Wilson-Fisher transition.

To decipher the critical properties of the quantum phase transition, we perform a finite-size scaling analysis to extract the critical exponents. The RG-invariant correlation-length ratios for different system sizes satisfy the following scaling form for sufficiently large $L$ and $J$ close to $J_c$:
\bea\label{scaling}
&&R_{\rm{EC}} = \mathscr{F}(L^{1/\nu}(J-J_c)),~~~
\eea
where $\nu$ is correlation-length exponent and $\mathscr{F}$ is a non-universal scaling function. Upon choosing the optimal value of $\nu$, the data points of relatively large $L$ should collapse on a single curve. The results of data collapse analysis are shown in \Fig{fig2}(a), giving rise to the correlation-length exponent $\nu=0.735\pm0.024$. Fixing $J$ at QCP $J_c=3.24$, we evaluate the correlation function of exciton order $C(r) = \avg{\phi_{\rm EC}(0)\phi_{\rm EC}(r)}$. The anomalous dimension of exciton order is determined by the scaling form $C(r) \sim r^{-1-\eta} $, the results of which are shown in \Fig{fig2}(b), yielding the anomalous dimension $\eta=0.051\pm0.036$. The obtained critical exponents $\nu=0.735\pm 0.024$ and $\eta=0.051\pm 0.036$ are reasonably consistent with previous results of 3D O(4) transition\cite{Kanaya1995PRD}, in which $\nu=0.7479(90)$ and $\eta = 0.0258(22)$. Hence, our large-scale QMC simulation provides convincing evidence that the quantum phase transition between the exciton insulator and SMG phase belongs to (2+1)D O(4) Wilson-Fisher universality class. 

\begin{figure}[tb]
\includegraphics[width=0.5\textwidth]{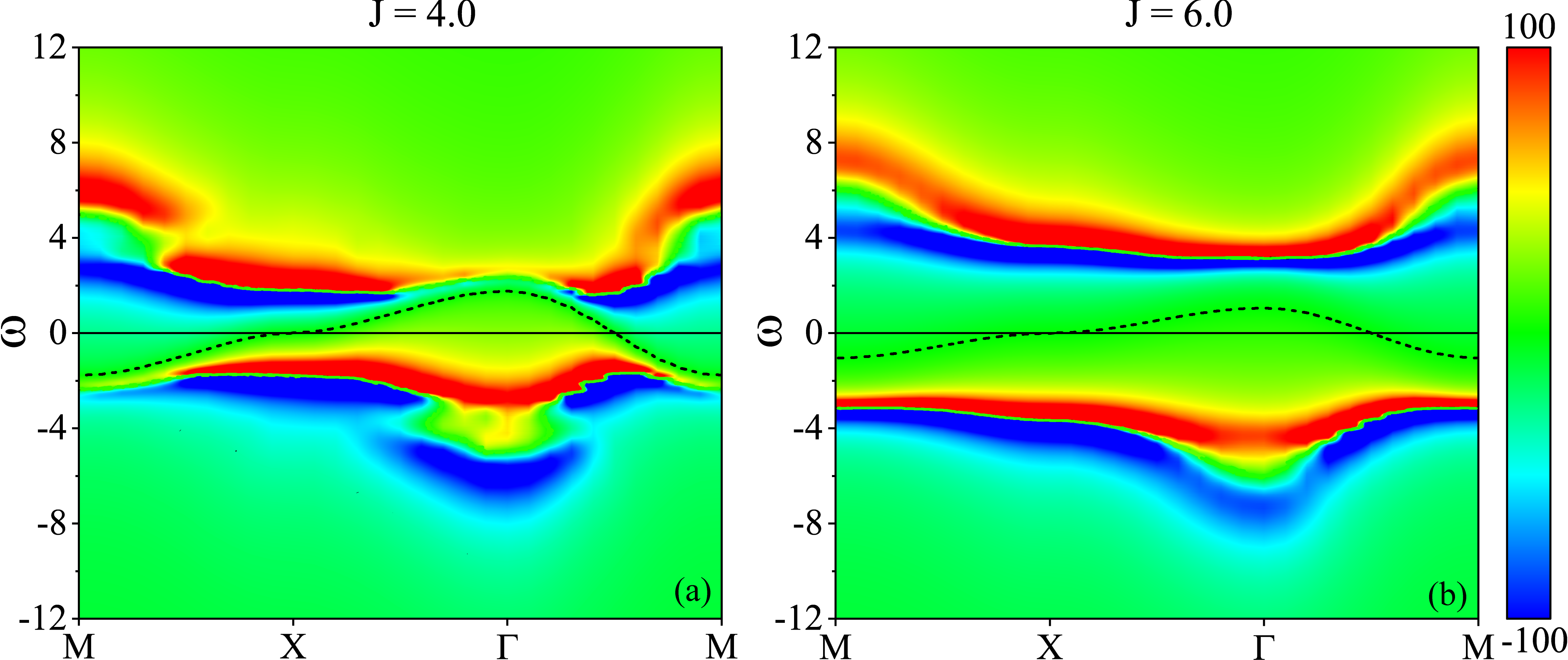}
\caption{ The real part of Green's function extracted from the spectral function by KK relation for $J=4.0$ (a) and $J=6.0$ (b). The dashed lines represent the zeros of Green's function. }
\label{fig4}
\end{figure}

{\it Spectral properties and Green's function zeros}: One prominent feature of the SMG insulator is the zeros of single-particle Green's function at low energy. The celebrated Luttinger's theorem dictates the volume enclosed by the poles of Green's function in momentum space yields the density of electrons. In the SMG state, the surface of Green's function poles is replaced by the zeros of Green's function, which defines Luttinger volume to match the particle density. To delve into such a prominent feature of the SMG phase hosted in \Eq{Hamiltonian}, we explore the spectral properties and extract the Green's function zeros in the model. 

Employing QMC simulation, we evaluate the imaginary-time single-particle Green's function and access the spectral function through stochastic analytical continuation (SAC)\cite{Sandvik1998SAC,Shao2023SAC}. The results of spectral function for different values of $J$ are presented in \Fig{fig3}. In the exciton insulator phase with weak interaction, for instance, $J=2.4$, the spectral function reveals the dispersion of a symmetry-breaking band insulator, displaying the minimum band gap located near the Fermi surface in the non-interacting limit. The dispersion exhibits the band-folding with momentum $(\pi,\pi)$, consistent with the ordering momentum of exciton order --- a clear indication of translational symmetry breaking. With increasing interaction strength $J$, exciton order is suppressed and the system evolves into a SMG phase. Deep in the SMG phase, the peaks of spectral function form two bands with relatively flat dispersion, resembling the spectrum of Mott insulator. Notably, with increasing $J$, the spectral peak in SMG phase exhibits remarkably pronounced broadening, manifesting an insulator triggered by strong interaction without well-defined quasi-particle.

Furthermore, we extract the real part of Green's function from the single-particle spectral function through the Kramers-Kronig (KK) relation:
\bea
\text{Re}\,G(k,\omega) = \frac{1}{2\pi} \int d\omega' \frac{A(k,\omega')}{\omega-\omega'}, 
\eea
and investigate the structures of Green's function zeros, defined by $G(k,\omega)=0$. The results for $J=4$ and $J=6$ are presented in \Fig{fig4}(a) and (b), respectively, in which the zeros of Green's function real part are dictated by the dashed line. The dispersion of Green's function's zeros appears and is reversed compared with the original non-interacting band dispersion. The ``bandwidth" of the Green's function zeros is suppressed with increasing interaction strength $J$ \cite{You2023arXiv}. Crucially, at $\omega=0$ the zeros of Green are precisely located in the Fermi surface of the model in the non-interacting limit. As a consequence, the volume enclosed by the surface of zeros is equivalent to the corresponding pole Fermi surface in the non-interacting limit, fulfilling the requirement of Luttinger theory.

{\it Relevance to $La_3 Ni_2 O_7$ superconductor}: As aforementioned, the model in \Eq{Hamiltonian} is potentially relevant to the high-$T_c$ superconductor $\rm{La}_3 \rm{Ni}_2 \rm{O}_7$, capturing the essential physics of Ni-$3d_{z^2}$ electrons in $\rm{La}_3 \rm{Ni}_2 \rm{O}_7$. As revealed by the recent ARPES measurements, the $3d_{z^2}$ electrons exhibit nearly flat dispersion and pronounced strong mass renormalization, in sharp contrast with the band dispersion of $3d_{x^2-y^2}$ electrons\cite{Zhou2023arXiv}. The nearly flat dispersion and large mass renormalization resemble the spectral function in the SMG insulating phase as shown in \Fig{fig3}, implying that the strong inter-layer spin interaction offers a possible origin of the electronic structure observed in $\rm{La}_3 \rm{Ni}_2 \rm{O}_7$\cite{Zhou2023arXiv}. The Fermi surface SMG phase is a plausible parent insulating state from which SC arises, similar to the AFM Mott insulating phase in cuprates.  More theoretical and experimental investigations in this direction are needed to corroborate this scenario.   

 Moreover, our unbiased QMC simulation suggests that it is feasible to extract the surface of Green's function zeros from the spectral function, albeit in the presence of the pronounced broadening of spectral peak and noise arising from analytical continuation. The numerical results shed new light on the observation of Fermi surface zeros in experiments. It is promising to implement the procedure in our study to extract the information of Green's function zeros from the spectral function measured in ARPES experiments and detect the experimental evidence of the Fermi surface SMG insulating phase. 

{\it Concluding remarks}: In this paper, we perform large-scale QMC simulation to investigate the SMG phase in an interacting bilayer model on a square lattice. The model features exciton/SC order breaking $\rm{SO}(4)$ pseudospin symmetry in the weak coupling regime and transitions to the SMG insulator in the presence of strong Heisenberg interaction. We determine the ground-state phase diagram and measure the critical properties of the quantum phase transition between the exciton insulator and SMG phase, which confirms an $\rm{O}(4)$ Wilson-Fisher universality class. We systematically study the spectral properties in the SMG insulator and exciton insulator. In the SMG insulator, the spectral peaks feature a relatively flat dispersion with pronounced broadening, resembling a Mott insulator without well-defined quasi-particles. More essentially, we determine the zeros of Green's function from the spectral function. The Green's function zeros in the zero frequency enclose the Fermi surface of the non-interacting band structure, replacing the roles of Green's function poles in the non-interacting limit. The Fermi volume enclosed by the surface associated with Green's function zeros is equivalent to Luttinger volume in the non-interacting limit, thus retaining Luttinger's theorem in the SMG insulating phase. The spectral properties revealed in the numerical simulation capture the prominent features of ARPES measurements of the Ni-$3d_{z^2}$ electrons in $\rm{La}_3 \rm{Ni}_2 \rm{O}_7$. Remarkably, the model \Eq{Hamiltonian} is still sign-problem-free away from half-filling, hence it is interesting to investigate the possible SC emerging from doping the SMG phase by unbiased QMC simulation, which is left for our future work.

\textit{Acknowledgement}: We are indebted to Kun Jiang, Yuan-Yao He and Yingfei Gu for stimulating discussions.  Z.X.L acknowledges support from the start-up grant of IOP-CAS. Y.Z.Y. is supported by the National Science Foundation Grant No. DMR-2238360. 

\bibliography{SMG}

% Custom numbering for equations, figures, and tables
\renewcommand{\theequation}{S\arabic{equation}}
\setcounter{equation}{0}
\renewcommand{\thefigure}{S\arabic{figure}}
\setcounter{figure}{0}
\renewcommand{\thetable}{S\arabic{table}}
\setcounter{table}{0}

\newpage
\begin{widetext}
\section{Supplemental Material for squarelattice-SMG}

\subsection{Section I: Projector determinant Quantum Monte-Carlo algorithm }
\label{sec:A2}

In this work, we implement projector Quantum Monte-Carlo (PQMC) simulation, which is an unbiased numerical algorithm for studying the ground state of interacting fermionic systems. The key scheme of PQMC is that one can access the accurate ground-state properties through imaginary-time evolution on a trial wave function, as long as the trial wave function is not orthogonal to the real ground-state wave function which is generically satisfied for a quantum many-body Hamiltonian. The ground-state wave function is accessed as: $| \psi_{G} \rangle = \lim_{\Theta \rightarrow \infty}e^{-\Theta \hat H}|\psi_{T} \rangle$, 
where $\Theta$ is the projective parameter, $| \psi_{G} \rangle$ is the ground state of Hamiltonian under consideration and $| \psi_{T} \rangle$ is trail wave function. For simplicity, the ground-state wave function of given non-interacting Hamiltonian is generally chosen as the trial wave function, which is a Slater-determinate. Under this scheme, the corresponding ground-state expectation value of observable is achieved as:   
\bea
\langle \hat O \rangle = \frac{\langle \psi_{G}|\hat O|\psi_{G} \rangle}{\langle \psi_{G}|\psi_{G} \rangle} = \frac{\langle \psi_{T}| e^{-\Theta \hat H} \hat O e^{-\Theta \hat H} | \psi_{T} \rangle}{\langle \psi_{T}| e^{-2\Theta \hat H} | \psi_{T} \rangle}
\label{PQMC}
\eea

In PQMC, one usually employs Trotter decomposition to discretize imaginary time and Hubbard-Stratonovich (H-S) transformation to decouple the interacting terms in the Hamiltonian. The Trotter decomposition is expressed as: 
\bea
e^{-\Theta \hat H} = \lim_{\Delta\tau \rightarrow 0} \[e^{-\Delta\tau \hat H_t} e^{-\Delta\tau \hat H_I} \]^{N\tau}
\eea
where $\Delta\tau = \frac{\Theta}{N_\tau}$ is the imaginary-time slice, $\hat H_t$ and $\hat H_I$ is the kinetic and interacting part of Hamiltonian, respectively. In our model, the kinetic and interacting parts of Hamiltonian are written as:
\bea
\hat{H} &=& \hat{H}_t + \hat{H}_I \\
\label{noninteracting}
\hat{H}_t &=& -t\sum_{\avg{ij}\sigma\alpha}(c^\dagger_{i\sigma\alpha}c_{j\sigma\alpha}+\mathrm{h.c.}) \\
\hat{H}_I &=& J\sum_{i} \vec{S}_{i,1}\cdot \vec{S}_{i,2}
\eea
where $c^\dagger_{i\sigma}$ creates an electron on site $i$ with spin polarization $\sigma=\uparrow,\downarrow$ and layer index $\alpha=1,2$. 
Here $t$ is the hopping amplitude of electrons between NN sites and $J$ denotes the strength of inter-layer AFM 
Heisenberg interaction. To facilitate the Hubbard-Strotonovich transformation on the interaction, the interacting term is rewritten as:
\bea
\hat H_I = \frac{J}{4}\sum_i \[ \( S_{i1}^x + S_{i2}^x \)^2 + \( S_{i1}^y + S_{i2}^y \)^2 + \( S_{i1}^z + S_{i2}^z \)^2 - \( S_{i1}^x - S_{i2}^x \)^2 - \( S_{i1}^y - S_{i2}^y \)^2 - \( S_{i1}^z - S_{i2}^z \)^2 \]
\eea
In our simulation, the discrete H-S transformation is implemented\cite{Assaadnote}:
\bea
e^{\Delta\tau \lambda A^2} = \sum_{l=\pm 1,\pm 2}\gamma\(l\) e^{\sqrt{\Delta\tau \lambda} \eta\(l\) A} + \mathcal{O}\(\Delta\tau^4\)
\eea
where the auxiliary field $\eta$ and $\gamma$ are taken the following values:
\bea
& \gamma \(\pm 1\) = 1 + \sqrt{6}/3 , \\
& \gamma\(\pm 2\) = 1 - \sqrt{6}/3 ,\\
& \eta\(\pm 1\) = \pm \sqrt{2\(3-\sqrt{6}\)}, \\
& \eta\(\pm 2\) = \pm \sqrt{2\(3+\sqrt{6}\)}.
\label{HS}
\eea
Under the transformation, the exponential of interacting term is expressed as: 
\bea
e^{-\Delta\tau \hat H_I} = 
 \prod_{{\sigma_1}=x,y,z} \[\sum_{a_1^{\sigma_1}=\pm 1,\pm 2}\gamma\(a_1^{\sigma_1}\) e^{\hat{h}_1^{\sigma_1}}\] \times \prod_{{\sigma_2}=x,y,z} \[\sum_{a_2^{\sigma_2}=\pm 1,\pm 2}\gamma\(a_2^{\sigma_2}\) e^{\hat h_2^{\sigma_2}}\]
\eea
where $\hat{h}_1^{\sigma_1} = i\sqrt{\Delta\tau J/4} \eta(a_1^{\sigma_1}) (S_{i1}^{\sigma_1} + S_{i2}^{\sigma_1})$, $\hat{h}_2^{\sigma_2} = \sqrt{\Delta\tau J/4} \eta(a_2^{\sigma_2}) (S_{i1}^{\sigma_2} - S_{i2}^{\sigma_2})$ and $a_1^{\sigma_1},a_2^{\sigma_2}=\pm1, \pm2$. Then, it is straightforward to employ the standard procedure of PQMC to evaluate the ground-state expectation value of observables as shown in \Eq{PQMC}\cite{Assaadnote,ZXLiQMCreview}. 

\begin{figure*}[tb]
\includegraphics[width=1.0\textwidth]{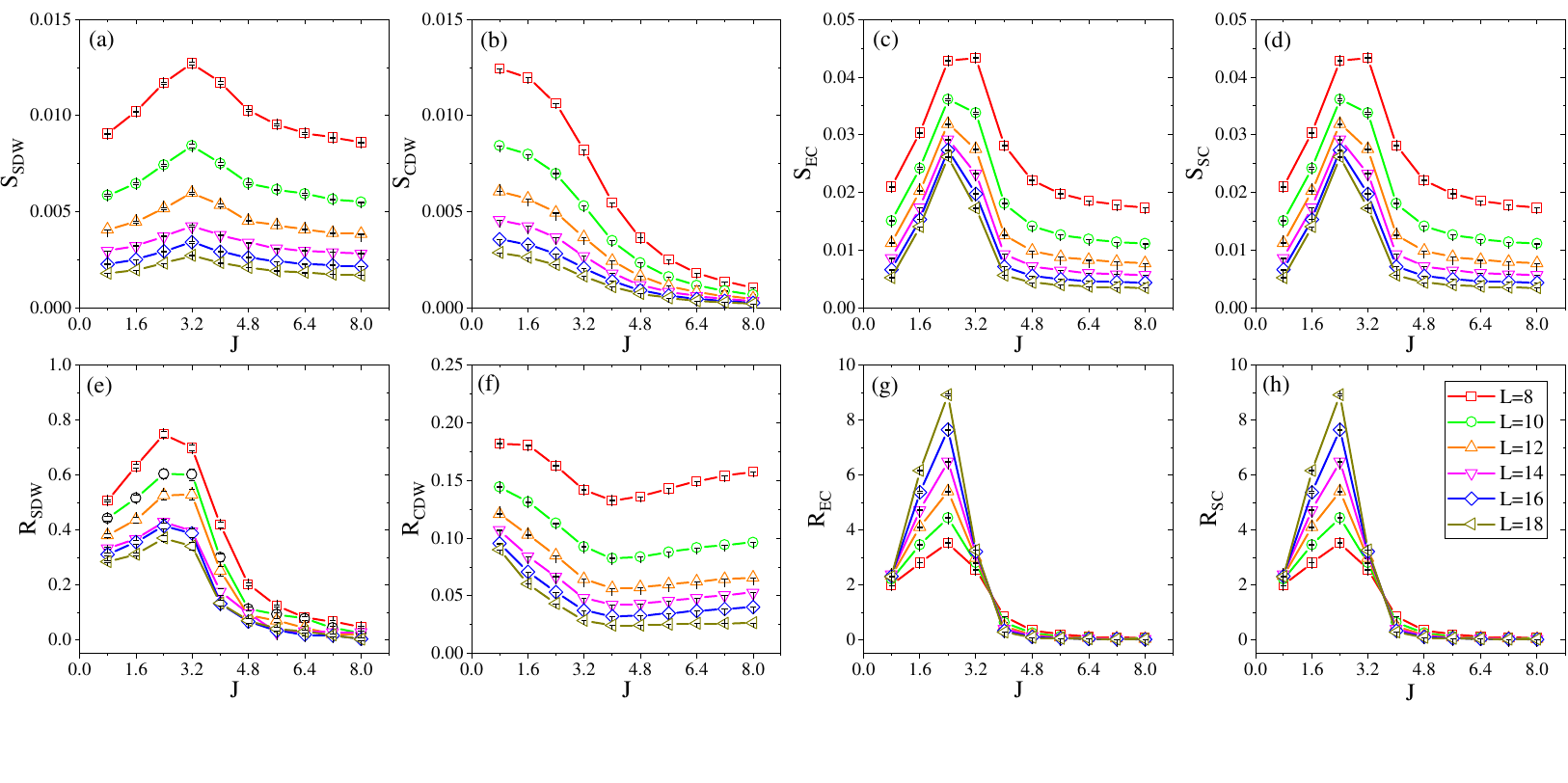}
\caption{(a)-(d) The results of structure factors for SDW, CDW, EC and SC. (e)-(h) The results of correlation-length ratios for SDW, CDW, EC and SC. The system sizes are $L=8,10,12,14,16,18$. }
\label{figs1}
\end{figure*}

\subsection{Section II: The absence of sign problem}
\label{sec:A1}

In this section, we demonstrate the absence of sign problem in the procedure of PQMC as illustrated in the last section. After H-S transformation, the decoupled interacting terms are expressed as:
\bea
\hat{h}_1^{\sigma} &=& i\sqrt{\Delta\tau J/4} \eta(a) (S_{i1}^{\sigma} + S_{i2}^{\sigma}) \\
\hat{h}_2^{\sigma} &=& \sqrt{\Delta\tau J/4} \eta(a) (S_{i1}^{\sigma} - S_{i2}^{\sigma})
\eea
where $\sigma = x,y,z$, $a = \pm 1,\pm2$ and $\vec{S}_{il} = (S^x_{il},S^y_{il},S^z_{il}) = \frac{1}{2}c^\dagger_{i\alpha l}\vec{\sigma}^{\alpha\beta}c_{i\beta l}$, with $\vec{\sigma} =(\sigma_x,\sigma_y,\sigma_z)$ being Pauli matrix. The definition of $\eta(a)$ is shown in \Eq{HS}. Remarkably, there exists a non-unitary transformation $\hat{T} = i \tau_x \sigma_y K$, where $\tau_x$ is the first Pauli matrix operating in the space of layer, $\sigma_y$ is the second Pauli matrix operating in the spin space, and $K$ is the operation of complex conjugation. It is straightforward to show that the non-interacting Hamiltonian and the interacting terms after H-S decoupling are invariant under such non-unitary transformation:
\bea
\hat{T} \hat{H}_t \hat{T}^{-1} &=& H_t \\
\hat{T} \hat{h}_1^\sigma \hat{T}^{-1} &=& \hat{h}_1^\sigma \\
\hat{T} \hat{h}_2^\sigma \hat{T}^{-1} &=& \hat{h}_2^\sigma 
\eea
where $\sigma=x,y,z$ and $\hat{H}_t$ is defined in \Eq{noninteracting}. Because $\hat{T}$ obeys $\hat{T}^2 = -1$, the model is sign-problem-free in QMC simulation owing to the principle introduced in \cite{CJWu2005PRB}.

\begin{figure*}[tb]
\includegraphics[width=0.75\textwidth]{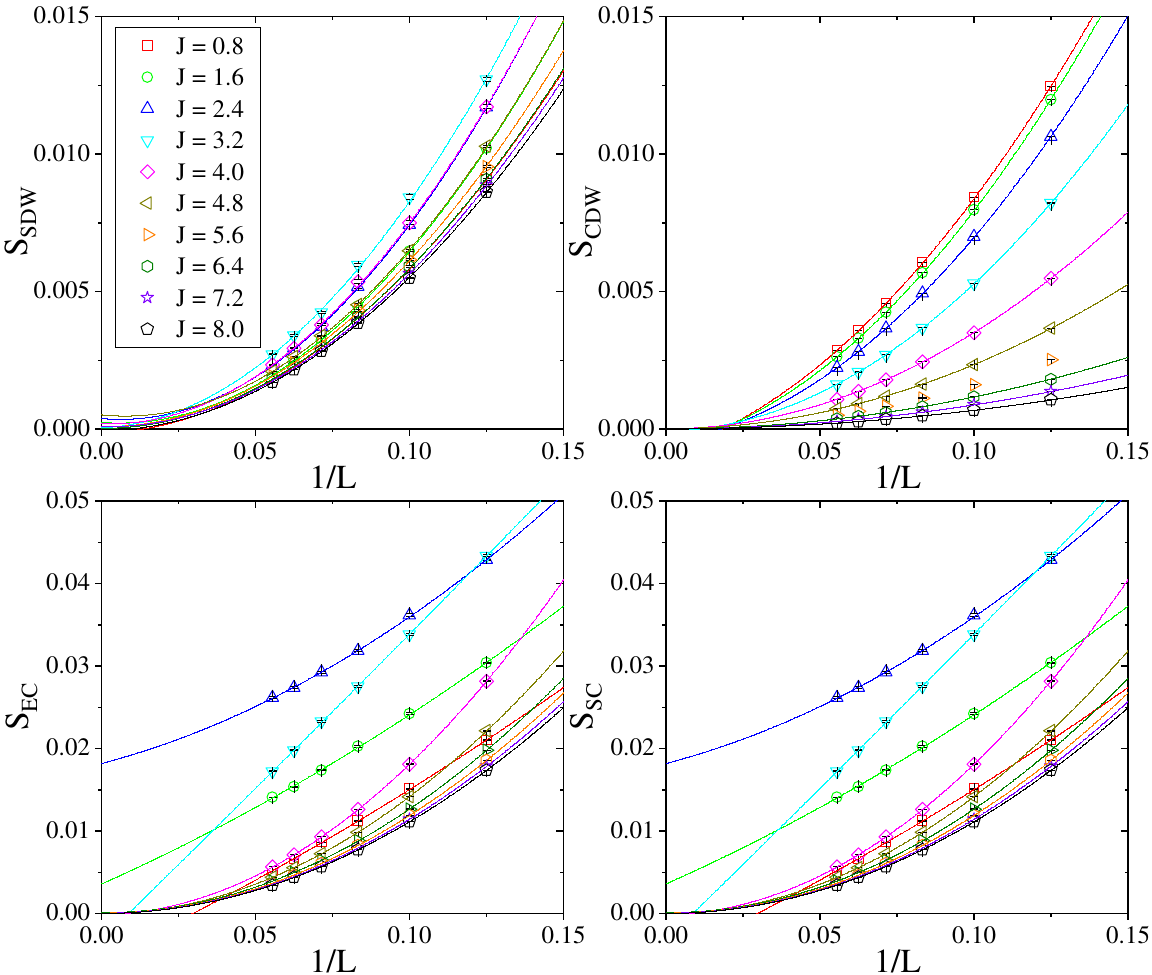}
\caption{ Finite-size scaling analysis of SDW (a), CDW (b), EC (c) and SC (d). The results are fitted by a second-order polynomial function of $1/L$. The system sizes are $L=8,10,12,14,16,18$. }
\label{figs2}
\end{figure*}

\subsection{III. The definition and results of structure factors and correlation-length ratios}

To characterize the symmetry spontaneously breaking phases and the phase transitions, we compute structure factors of various symmetry breaking orders, which are defined as following:
\bea
S(\vec{Q},L) = \frac{1}{L^4} \sum_{ij} \avg{ \hat{O}(i)\hat{O}(j)} e^{i \vec{Q}\cdot (\vec{r}_i - \vec{r}_j)}
\eea
where $\hat{O}(i)$ is the symmetry breaking operator field located at site $i$, $L$ is the linear system size and $\vec{Q}$ is the momentum of symmetry breaking order under consideration. In our cases, we consider the spin-density wave (SDW) and charge-density wave (CDW) orders. The corresponding operator fields are defined as:  
\bea
O_{\rm{SDW}}(i) = \sum_{\alpha} c^\dagger_{i\uparrow\alpha}c_{i\uparrow\alpha}-c^\dagger_{i\downarrow\alpha}c_{i\downarrow\alpha}  \quad O_{\rm{CDW}}(i) = \sum_{\alpha} c^\dagger_{i\uparrow\alpha}c_{i\uparrow\alpha}+c^\dagger_{i\downarrow\alpha}c_{i\downarrow\alpha}
\eea
where $\alpha$ denotes the index of layer, $\uparrow$/$\downarrow$ is spin polarization. Here we only consider spin order in z-direction without the loss of generality because the model in Hamiltonian \Eq{Hamiltonian} preserves spin $SU(2)$ symmetry. For the bilayer model, the interlayer exciton condensation (EC) order is also considered in our study. The operator field is defined as:
\bea
O_{\rm{EC}}(i) = \sum_\sigma c^\dagger_{i\sigma 1}c_{i\sigma 2}
\eea
where $\sigma = \uparrow/\downarrow$ is the spin polarization and $1/2$ is the layer index. Additionally, we consider the order of superconductivity (SC). In our case, the dominant SC channel is the interlayer on-site spin singlet pairing, with the corresponding operator field defined as:
\bea
O_{\rm{SC}}(i) = c^\dagger_{i\uparrow 1} c^\dagger_{i\downarrow 2} - c^\dagger_{i\downarrow 1} c^\dagger_{i\uparrow 2}
\eea
For SC ordering, the dominant ordering momentum is $(0,0)$.  For the order parameter field involving particle-hole instabilities, including SDW, CDW and EC, the dominant ordering carries momentum $(\pi,\pi)$ in our cases owing to the Fermi surface nesting of non-interacting band structure.  We also confirm them via scrutinizing the momentum distribution of structure factors for various orders. 

To reveal the possible SSB in model \Eq{Hamiltonian}, we evaluate the structure factors at peaked momentum for different system sizes, as presented in \Fig{figs1}(a)-(d). In the weakly interacting regime, the structure factors for EC and SC orders are obviously larger than the ones for SDW and CDW. The finite-size scaling of structure factors for SDW, CDW, EC and SC are shown in \Fig{figs2}. The results clearly show that SDW and CDW long-ranged order is absent in the entire interaction regime of the model \Eq{Hamiltonian}. The EC and inter-layer singlet SC order exist in the weakly interacting regime and vanish when interaction is strong. The degeneracy of EC and SC order is also explicitly witnessed by the equivalence of structure factors.   

\begin{figure*}[tb]
\includegraphics[width=0.75\textwidth]{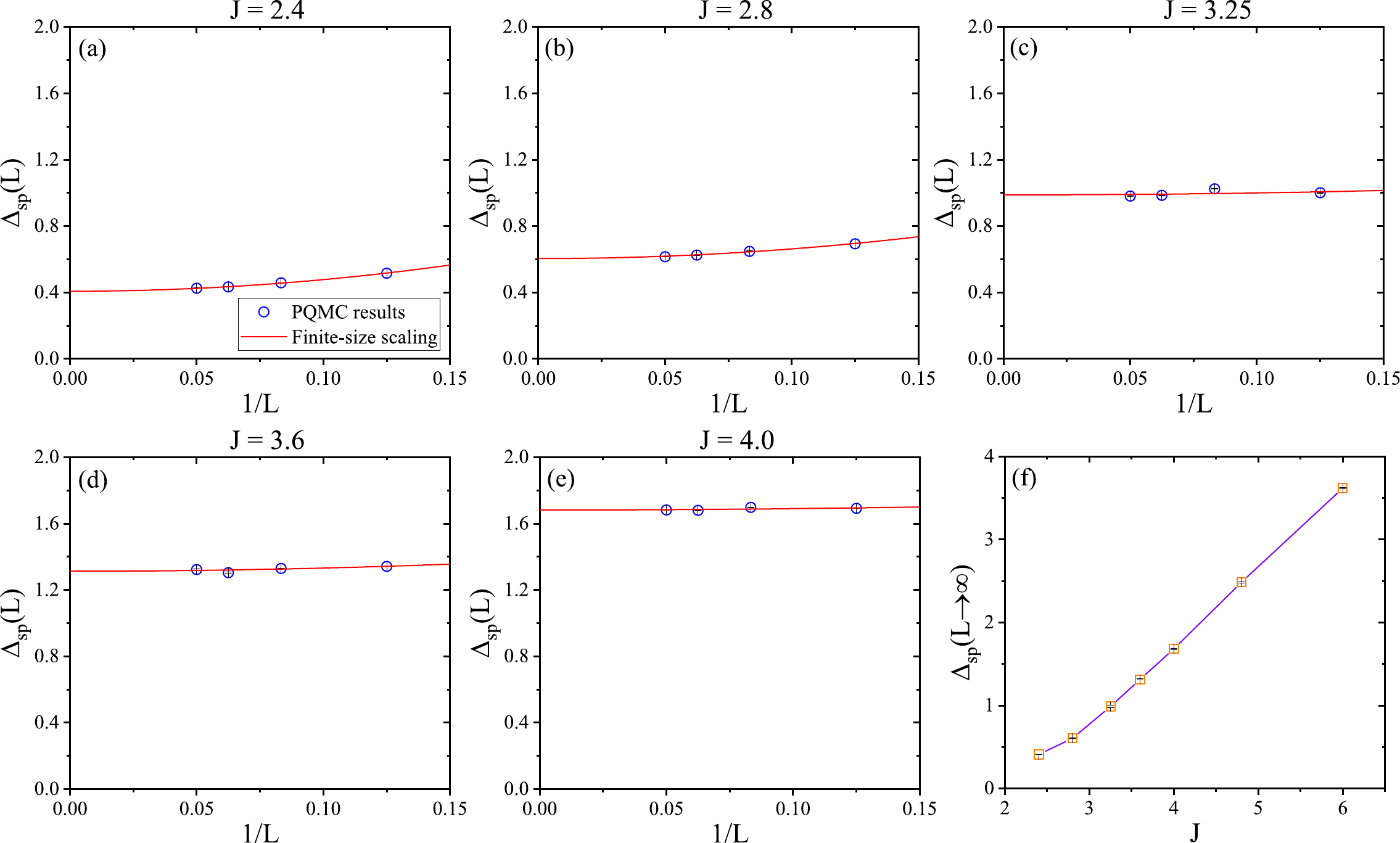}
\caption{ (a)-(e) Finite-size scaling of single-particle gap $\Delta_{\rm sp}(L)$ for various interaction strength $J$. The $\Delta_{\rm sp}$ is extracted at the momentum $(\frac{\pi}{2},\frac{\pi}{2})$, at which the gap size is minimum. (f) The results of single-particle gap at thermodynamic limit obtained by finite-size scaling as a function of $J$.   }
\label{figs3}
\end{figure*}

To further verify the conclusions, we evaluate the correlation-length ratio for different system sizes, which is believed to bear weaker finite-size effects compared with the direct extrapolation of structure factors. The correlation-length ratio is defined as:
\bea
R(L) = \frac{S(\vec{Q}_m,L)}{S(\vec{Q}_m+\delta\vec{q},L)} -1
\eea
where $S(\vec{Q}_m,L)$ is the structure factor of the order under consideration,  $\vec{Q}_m$ is the ordering momentum, and $\delta\vec{q} = (\frac{2\pi}{L},\frac{2\pi}{L})$ denotes the minimum momentum shift from ordering momentum $\vec{Q}_m$ on a lattice with linear system size $L$.  For the disordered phase of the order parameter under consideration, $R(L)$ decreases with $L$ and $R(L) \rightarrow 0$ in the thermodynamic limit, whereas for the long-ranged ordered phase, $R(L)$ increases with $L$ and $R(L) \rightarrow \infty$ in the thermodynamic limit. At  putative quantum critical point, RG-invariant quantity $R(L)$ is independent on system size as $L$ is sufficiently large. The results of correlation-length ratios for SDW, CDW, EC and SC are presented in \Fig{figs1}(e)-(h). For SDW and CDW order, $R(L)$ decreases with system size in the whole interacting regime. For EC and SC order, a quantum phase transition from long-ranged ordered phase to the disordered phase occurs with increasing interaction strength. The degeneracy between EC and SC order is further verified by the equivalence of correlation-length ratios for two orders.

\subsection{IV. The finite-size scaling of single-particle gap}
To reveal the nature of SMG insulating phase, a prerequisite is the single-particle excitation is gaped. In this section, we present the finite-size scaling results of single-particle gap. In QMC simulation, single-particle gap is accessible by evaluating the time-dependent single-particle Green's function $G(\vec{k},\tau) = \avg{c_{\vec k}(0)c^\dagger_{\vec k}(\tau)}$, which exhibits the scaling behavior $G(\vec{k},\tau) \sim e^{-\Delta_{sp}(\vec{k})\tau}$ as imaginary-time $\tau$ is long enough. Hence, the single-particle gap at momentum $\vec k$ is extracted from the scaling of $G(\vec k,\tau)$ versus $\tau$. According to the numerical results of single-particle gap in the Brillouin zoom, it is shown that the single-particle gap is minimum at momentum $(\pm \frac{\pi}{2},\pm \frac{\pi}{2})$. Fixing $\vec{k}=(\frac{\pi}{2},\frac{\pi}{2})$, we compute time-dependent Green's function for different system sizes and various $J$, and extract the single-particle gap from the scaling behavior of $G(\vec{k},\tau)$. The results of single-particle gap are included in \Fig{figs3}. In the entire interaction strength regime we consider, the single-particle gap at thermodynamic limit is finite, indicating that the ground state is an insulator. Particularly, in the strongly interacting regime $J>3.24$ where the ground state does not involve any spontaneous symmetry breaking, the single-particle exciton features pronounced gap, demonstrating the nature of a SMG insulating phase.

\end{widetext}

\end{document}